\documentclass[12pt]{article}
\usepackage{amssymb,amsmath,epsfig}

\begin{document}

\title{\bf Expanding and Collapsing Scalar Field Thin Shell}
\author{M. Sharif \thanks{msharif.math@pu.edu.pk} and G. Abbas
\thanks{abbasg91@yahoo.com}\\
Department of Mathematics, University of the Punjab,\\
Quaid-e-Azam Campus, Lahore-54590, Pakistan.}

\date{}
\maketitle
\begin{abstract}
This paper deals with the dynamics of scalar field thin shell in the
Reissner-Nordstr$\ddot{o}$m geometry. The Israel junction conditions
between Reissner-Nordstr$\ddot{o}$m spacetimes are derived, which
lead to the equation of motion of scalar field shell and
Klien-Gordon equation. These equations are solved numerically by
taking scalar field model with the quadratic scalar potential. It is
found that solution represents the expanding and collapsing scalar
field shell. For the better understanding of this problem, we
investigate the case of massless scalar field (by taking the scalar
field potential zero). Also, we evaluate the scalar field potential
when $p$ is an explicit function of $R$. We conclude that both
massless as well as massive scalar field shell can expand to
infinity at constant rate or collapse to zero size forming a
curvature singularity or bounce under suitable conditions.
\end{abstract}

\section{Introduction}

The idea of the geon (electromagnetic-gravitational entity) by
Wheeler \emph{et al.} \cite{1,2} leads to investigation of scalar
field in general relativity (GR). In GR, the scalar field appears in
the low energy limit of string theory \cite{1a}. Still, there is no
observational evidence about the existence of such particles that
are associated with the scalar field. However, the study of
gravitational collapse of compact objects in the scalar-tensor
theories imply that scalar field might be the source of scalar
gravitational waves that can be detected by the advanced detectors
\cite{2a}. Further, the observational facts from the binary pulsar
may include or exclude the presence of scalar field in GR \cite{3a}.
In astrophysical context, boson stars are such compact objects that
are composed of scalar field. Probably, such stars were created in
the early universe as Higgs particles condensate \cite{4a}.
Recently, it has been proposed that boson stars are the candidates
for dark matter \cite{5a}. Also, some remarkable similarities are
suggested between neutron and boson stars.

Dynamics of scalar fields has been the subject of particular
interest in both cosmological as well as astrophysical situations.
In cosmological scenarios, scalar fields have a great attraction
because such fields act as an effective cosmological constant in
deriving the inflation. The nature of spacetime singularity for
massless scalar field has been investigated in recent years on
spherical collapse models \cite{3}-\cite{10}. Bhattacharya \emph{et
al.} \cite{11} discussed the gravitational collapse of a minimally
coupled massless scalar field and examined the possibility of
existence of nonsingular models where collapse could be freezed
under suitable conditions. All these studies investigate the
dynamics of massless scalar field using the exact solutions of the
field equations. Kaup \cite{4a} was among the pioneers to study
complex massive scalar field configuration. Ruffini and Bonazzola
\cite{13} explored spherically symmetric system and determined the
equilibrium conditions for boson stars solutions.

There is another formalism to study the dynamics of the matter field
referred to as "\textit{thin shell formalism}" developed by Israel
\cite{14}. This is one of the exactly solvable formulation in GR
which is widely used to understand gravitational collapse and other
cosmological as well as astrophysical processes. It involves a
discontinuity of the extrinsic curvatures of the interior and
exterior regions across a boundary surface. The jump in the
extrinsic curvature across the boundary surface is caused by a thin
layer of matter. In this formulation, the set of equations leads to
equations of motion (much simpler) corresponding to the field
equations, involving curvature on one side and matter on the other
side. The solution of these equations provides full understanding of
the dynamics of system. In the relativistic astrophysics, the thin
shell equations help to study the properties of the compact objects.

Pereira and Wang \cite{15} studied gravitational collapse of the
cylindrical shell made of counter rotating dust particles by using
the Israel thin shell formalism. Sharif \emph{et al.}
\cite{16}-\cite{18} have investigated plane and spherically
symmetric gravitational collapse by using this formulation. This
approach was generalized to thin charged shell without pressure by
De La Cruz and Israel \cite{19}. Kuchar \cite{20} and Chase
\cite{21} treated the charged thin shell problem with pressure by
using polytropic equation of state. There are a number of papers
devoted to handle the charged thin shell problems. Boulware
\cite{22} studied the time evolution of the charged thin shell and
showed that their collapse could form a naked singularity if and
only if density is negative. Farrugia and Hajicek \cite{23}
investigated third law of black hole mechanics in the Reissner
Nordstr$\ddot{o}$m (RN) geometry. N\'{u}\~{n}ez \cite{24} studied
the oscillating perfect fluid shell. Also, N\'{u}\~{n}ez \emph{et
al.} \cite{25} explored the stability and dynamical behavior of the
real scalar field for the Schwarzschild geometry in single null
coordinate.

The purpose of the present paper is to study the application of
Israel thin shell formalism for the scalar fields in the context of
GR. In particular, we study the dynamical behavior of scalar field
thin shell in charged background. To this end, we take interior and
exterior regions as RN solutions, the Israel thin shell formalism is
used to derive equations of motion of the shell for the perfect
fluid. We then specify that the perfect fluid is generated by scalar
field which leads to scalar field equations of motion. The solution
of these equations provide expanding, collapsing or oscillating
scalar field shell. The plan of the paper is as follows: Equations
of motion for scalar field model with quadratic potential and their
physical interpretation are presented in section \textbf{2}.
Sections \textbf{3} and \textbf{4} deal with dynamics of the
massless and massive scalar field with arbitrary scalar potential,
respectively. We summarize our results in the last section.

\section{Dynamical Equations}

In this section, we use Israel thin shell formulation to derive
equations of motion of the scalar field shell. For this purpose, we
take a $3D$ spacelike boundary surface ${\Sigma}$, which splits the
two $4D$ spherically symmetric spacetimes $V^+$ and $V^-$. The
interior and exterior regions $V^-$and~$V^+$, respectively are
described by the RN metrics given by
\begin{equation}\label{1}
(ds)_\pm^2={\eta}_\pm dT^2-\frac{1}{\eta}_\pm
dR^2-R^2(d\theta^2+\sin^2{\theta}d\phi^2),
\end{equation}
where $\eta_\pm(R)=1-\frac{2M_\pm}{R}+\frac{Q_\pm^2}{R^2}$, $M_\pm$
and $Q_\pm$ are the mass and charge, respectively. The subscripts
$+$ and $-$ represent quantities in exterior and interior regions to
${\Sigma}$, respectively. Further, it is assumed that charge in both
regions is the same, i.e., $Q_-= Q_+ =Q $. The strength of the
electric field on the shell can be described by the Maxwell field
tensor, $F_{TR}=\frac{Q}{R^2}=-F^{RT}$.

The energy-momentum tensor of the electromagnetic field is
\begin{equation}\label{2}
{T_{\delta}^{\nu}}^{(em)}=\frac{1}{4{\pi}}
(-F^{{\nu}{\lambda}}F_{{\delta}{\lambda}}+\frac{1}{4}\delta^{\nu}_{\delta}
F_{{\pi}{\lambda}}F^{{\pi}{\lambda}}).
\end{equation}
By employing the intrinsic coordinates $(\tau,\theta,\phi)$ on
${\Sigma}$ at $R=R(\tau)$, the metrics (\ref{1}) on ${\Sigma}$
become
\begin{equation}\label{3}
{(ds)_\pm^2}_\Sigma=[\eta(R)-\frac{1}{\eta(R)}
(\frac{dR}{dT})^2]dT^2-R^2(\tau)(d\theta^2+\sin^2{\theta}d\phi^2).
\end{equation}
Here $T$ is also a function of $\tau$ and it is assumed that
$g_{00}>0$, so that $T$ is a timelike coordinate. Also, the induced
metric on the boundary surface ${\Sigma}$ is given by
\begin{equation}\label{4}
{(ds)^2}_\Sigma=d\tau^2-a^2(\tau)(d\theta^2+\sin^2{\theta}d\phi^2).
\end{equation}
The continuity of the first fundamental form gives
\begin{equation}\label{5}
[\eta(R_\Sigma)-\frac{1}{\eta(R_\Sigma)}
(\frac{dR_\Sigma}{dT})^2]^{\frac{1}{2}}dT=(d\tau)_\Sigma,\quad
R(\tau)=a(\tau)_{\Sigma}.
\end{equation}
Now the unit normal ${n_\mu}^\pm$ to ${\Sigma}$ in $V^{\pm}$
coordinates can be evaluated as
\begin{eqnarray}\label{6}
{n_\mu}^\pm=(-\dot{R}(\tau),\dot{T},0,0),
\end{eqnarray}
where dot represents differentiation with respect to $\tau$.

The non-vanishing components of the extrinsic curvature are
\begin{eqnarray}\label{8}
K^\pm_{\tau\tau}=\frac{d}{dR}\sqrt{\dot{R}^2+\eta_\pm},\quad
K^\pm_{\theta\theta}=-R\sqrt{\dot{R}^2+\eta_\pm},\quad
K^\pm_{\phi\phi}=K^\pm_{\theta\theta}\sin^2{\theta}.
\end{eqnarray}
The surface energy-momentum tensor is defined by
\begin{equation}\label{9}
S_{ij}=\frac{1}{\kappa}\{[K_{ij}]-\gamma_{ij}[K]\},
\end{equation}
where ${\kappa}$ is a coupling constant, $\gamma_{ij}$ is induced
metric on $\Sigma$ and
\begin{equation}\label{10}
[K_{ij}]=K^+_{ij}-K^-_{ij},\quad [K]=\gamma^{ij}[K_{ij}].
\end{equation}
The surface energy-momentum tensor for a perfect fluid is
\begin{equation}\label{11}
S_{ij}=(\rho+p){u_i}{u_j}-p\gamma_{ij},
\end{equation}
where $u_i=\delta^{0}_{{i}}$. Using Eqs.(\ref{5}), (\ref{9}) and
(\ref{11}), we can find
\begin{equation}\label{13}
\rho=\frac{2}{\kappa R^2}[K_{\theta\theta}],\quad
p=\frac{1}{\kappa}\{K_{tt}-\frac{[K_{\theta\theta}]}{R^2}\}.
\end{equation}

Inserting the non-zero components of the extrinsic curvature
components, we get
\begin{eqnarray}\label{14}
(\zeta_+-\zeta_{-})+\frac{\kappa}{2}\rho R=0,\\\label{15}
\frac{d}{dR}(\zeta_+-\zeta_{-})+\frac{1}{R}(\zeta_+-\zeta_{-})-{\kappa}p=0,
\end{eqnarray}
where $\zeta_\pm=\sqrt{\dot{R}^2+\eta_\pm}$. Making use of
Eq.(\ref{14}) in (\ref{15}), it follows that
\begin{equation} \label{16}
\dot{m}+ p\dot{A}=0,
\end{equation}
where $m(=\rho A )$ and $A(=4\pi R^2(\tau))$ stand for the
integrated total energy density at some time and area of the shell,
respectively. The conservation of surface energy-momentum tensor
leads to the same equation as Eq.(\ref{16}) and hence this equation
is known as energy conservation law on the shell. It is mentioned
here that this equation can be solved by using the equation of state
$p=k\rho$, and its solution is
\begin{eqnarray}\label{16a}
\rho=\rho_0(\frac{R_0}{R})^{2(k+1)},
\end{eqnarray}
where $R_0$ is the position of the shell at $t=t_0$ and $\rho_0$ is
the density of matter on the shell at position $R_0$. Using this
value of $\rho$ in the definition of $m$, we obtain
\begin{eqnarray}\label{16b}
m=4\pi\rho_0\frac{R_0^{(2k+2)}}{{R^{2k}}}.
\end{eqnarray}
From Eq.(\ref{14}), equation of motion of the shell is given by
\begin{equation}\label{17}
\dot{R}^2+V_{eff}(R)=0,
\end{equation}
where the effective potential $V_{eff}(R)$ is
\begin{equation}\label{18}
V_{eff}(R)=1-\left(\frac{M_+ -M_{-}}{m}\right)^2+\left(\frac{Q}
{R}\right)^2-\frac{(M_+ + M_{-})}{R} -\left(\frac{m}{2 R}\right)^2.
\end{equation}
Notice that we have used $\kappa=8\pi$ to derive this equation.

To see the effects of charge parameter $Q$ on the dynamics of shell,
we re-write Eq.(\ref{17}) by using the above equation as follows
\begin{equation}\label{17b}
\dot{R}=\pm\sqrt{\left(\frac{M_+ -M_{-}}{m}\right)^2-\left(\frac{Q}
{R}\right)^2+\frac{(M_+ + M_{-})}{R} +\left(\frac{m}{2
R}\right)^2-1.}
\end{equation}
Here $+(-$) correspond to expansion (collapse) of shell and $m$ is
the same as defined after Eq.(\ref{16}). The term $\frac{Q^2}{R^2}$
(Coulomb repulsive force) in $\dot{R}$ (velocity of shell with
respect to stationary observer) indicates that charge reduces
velocity of the shell with respect to stationary observer. This
velocity also depends on position of the observer, whether the
observer is located inside or outside the shell. Further,
Eq.(\ref{18}) implies that charge parameter increases the effective
potential $V_{eff}$. In coming sections, we shall see that
throughout the dynamics of shell (composed of either massless or
massive scalar field), charge parameter reduces and increases
velocity of the shell with respect to stationary observer and
$V_{eff}$, respectively. Thus initially the velocity of the shell
with respect to stationary observer in the RN background is slower
as compared to the Schwarzschild case (as shown in Figure
\textbf{1}). We conclude that electrostatic repulsive force in RN
background tries to balance with the gravitational force due to the
shell and hence the shell velocity with respect to stationary
observer is slow in this case as compared to uncharged case.
\begin{figure}
\center\epsfig{file=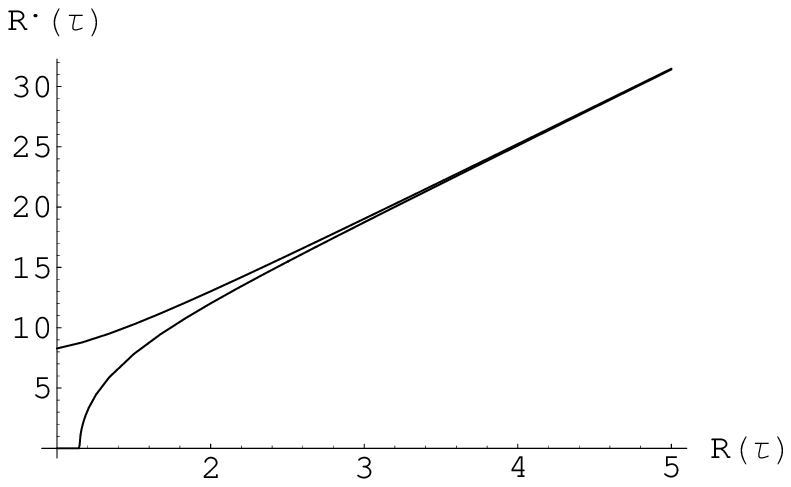, width=0.45\linewidth} \caption{Behavior
of the shell velocity with respect to stationary observer, when
$M_+=1,~M_-=0$, $k=\rho_0=R_0=1$ and $Q=1$. The upper and lower
curves correspond to uncharged and charged cases, respectively. It
is clear that initially velocity in the charged case is less than
the uncharged case. Velocities in both cases match for larger values
$R$, as term $\frac{Q^2}{R^2}$ becomes negligible for larger values
$R$} \center\epsfig{file=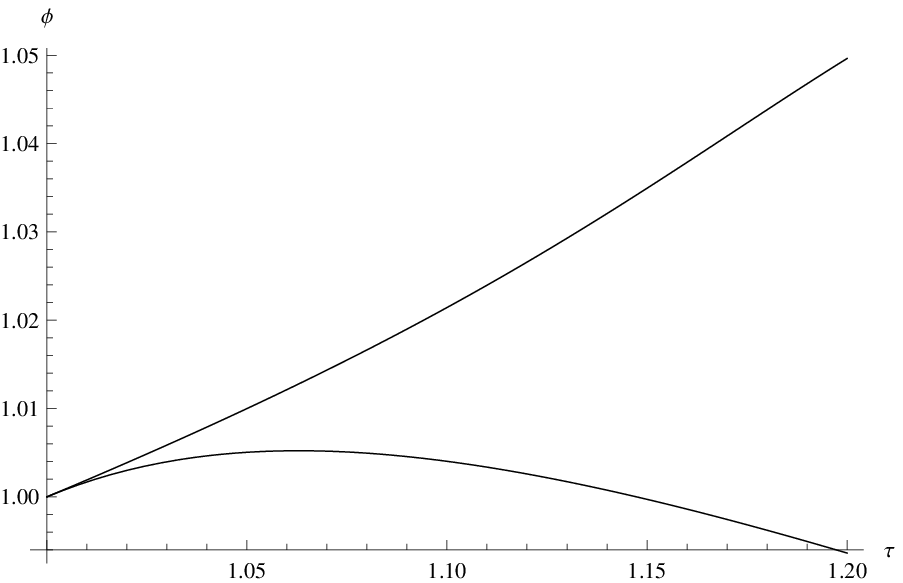,
width=0.45\linewidth}\epsfig{file=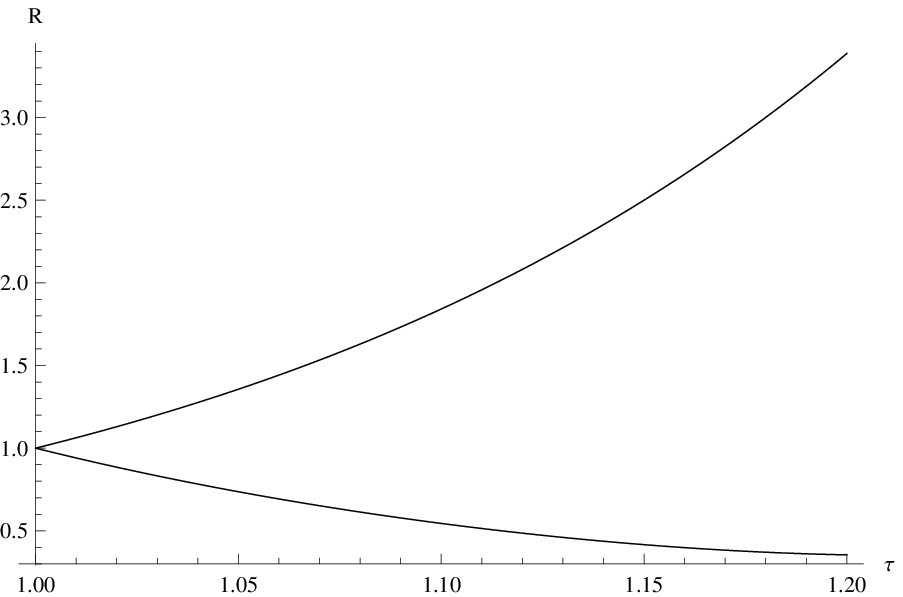, width=0.45\linewidth}
\caption{The left graph is the behavior of the shell radius $R$
while the right graph is the behavior of scalar field. Both these
graphs have been plotted by using $M_+=1,~\tilde{
m}=1,~M_-=0,~Q=1$,~$\dot{\phi}(1)=0.19$ and $\phi(1)=R(1)=1$.}
\end{figure}

In order to study the dynamics of scalar field shell, we specify
that the perfect fluid is generated by a scalar field. The energy
density and the pressure of the scalar field can be written as
\begin{eqnarray}\label{19}
\rho=\frac{1}{2}[{\phi}_{,~\nu }{\phi}^{,~\nu }+2V(\phi)], \quad
p=\frac{1}{2}[{\phi}_{,~\nu }{\phi}^{,~\nu }-2V(\phi)],
\end{eqnarray}
where $V(\phi)=\tilde{m}^2\phi^2$ is the potential term which
contributes to provide mass of the scalar field. We note that the
scalar field will be massless in the absence of such term. Using
Eq.(\ref{19}), we can write the energy-momentum tensor of the scalar
field as follows
\begin{equation}\label{22}
S_{ij}={\nabla}_i{\phi}{\nabla}_j{\phi}-\gamma_{ij}~[\frac{1}{2}({\nabla}{\phi})^2-V({\phi})].
\end{equation}
Since the induced metric (4) depends only on $\tau$, so $\phi$ also
depends on $\tau$. Thus Eq.(\ref{19}) leads to
\begin{equation}\label{23}
\rho= \frac{1}{2}[{\dot{\phi}^2}+2V(\phi)],\quad
p=\frac{1}{2}[{\dot{\phi}^2}-2V(\phi)].
\end{equation}
In terms of the scalar field, the integrated total energy density of
the shell at some time is
\begin{equation}\label{24}
m=2{\pi}R^2[{\dot{\phi}^2}+2V(\phi)].
\end{equation}
Using Eqs.(\ref{23}) and (\ref{24}) in Eq.(\ref{16}), we get
\begin{equation}\label{25}
\ddot{\phi}+\frac{2\dot{R}}{R}\dot{\phi}+\frac{{{\partial}V}}{{{\partial}{\phi}}}=0.
\end{equation}
This is the Klien-Gordon (KG) equation,
$\square\phi+\frac{{{\partial}V}}{{\partial}{\phi}}=0$, in
coordinate system of the shell metric (4). In terms of the scalar
field, the effective potential is
\begin{eqnarray}\label{26}
V_{eff}(R)&=&1-\left(\frac{M_+
-M_{-}}{2{\pi}R^2({\dot{\phi}^2}+2V(\phi))}
\right)^2+\left(\frac{Q}{ R}\right)^2-\frac{(M_+ +M_{-})}{R}\nonumber\\
&-&[{\pi}R({\dot{\phi}^2}+2V(\phi))]^2.
\end{eqnarray}

Now we solve the KG equation (\ref{25}) and equation of motion
(\ref{17}) (with Eq.(\ref{26})) simultaneously for $\phi(\tau)$ and
$R(\tau)$. In this case, the exact solution is not possible. We
solve these equations numerically by assuming the following initial
conditions: $\dot{\phi}(1)=0.19$ and $\phi(1)=R(1)=1$. The graphs of
these equations for the set of initial data are shown in Figure
\textbf{2}. The left graph shows the behavior of the shell radius
$R$ in which upper and lower curves represent the expanding and
collapsing shell, respectively. The right graph is the behavior of
scalar field whose upper and lower curves represent the collapsing
and expanding shell, respectively. In case of collapse (upper
curve), scalar field density $\phi$ goes on increasing while in case
of expansion (lower curve), this comes to a point on $\tau$-axis
implying that scalar field decays to zero value in this case.

\section{Massless Scalar Field}

A scalar field becomes massless, when scalar potential, $V(\phi)$,
is zero. In this case, the KG equation reduces to
$\ddot{\phi}+\frac{2\dot{R}}{R}\dot{\phi}=0$ whose solution is
$\dot{\phi}=\frac{\Omega}{R^2}$, where $\Omega$ is an integration
constant. Thus the equation of motion (\ref{17}) with Eq.(\ref{26})
takes the form
\begin{equation}\label{32}
\dot{R}^2+1-\left(\frac{M_+ -M_{-}}{2{\pi}{\Omega}^2}
\right)^2R^4+\left(\frac{Q}{ R}\right)^2-\frac{(M_+ + M_{-})}{R}-
\frac{\pi^2{\Omega}^4}{R^{6}}=0.
\end{equation}
\begin{figure}
\center\epsfig{file=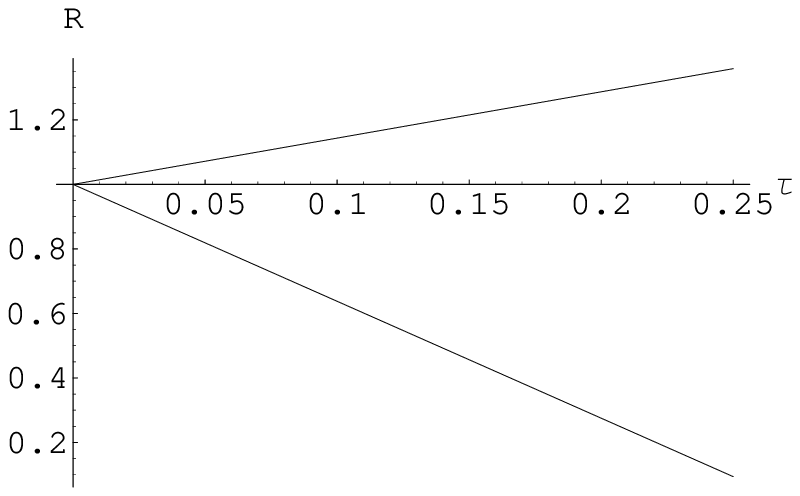, width=0.45\linewidth}\epsfig{file=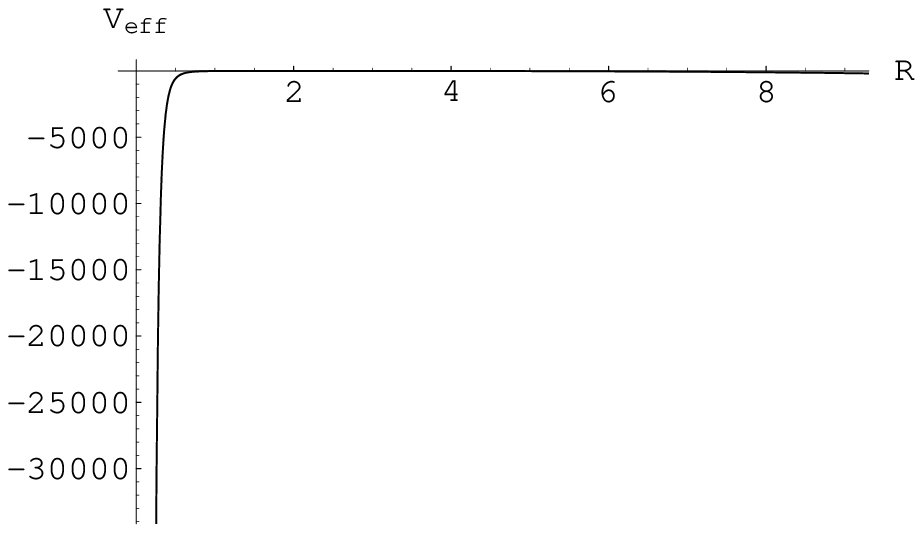,
width=0.35\linewidth} \caption{The left graph shows the shell radius
for massless scalar field case (Eq.(\ref{32})). The right graph is
the effective potential for massless scalar field (Eq.(\ref{34}))
with $\Omega=1$, keeping all the remaining parameters and initial
conditions fixed as in Figures 1, 2.}
\end{figure}
\begin{figure}
\center\epsfig{file=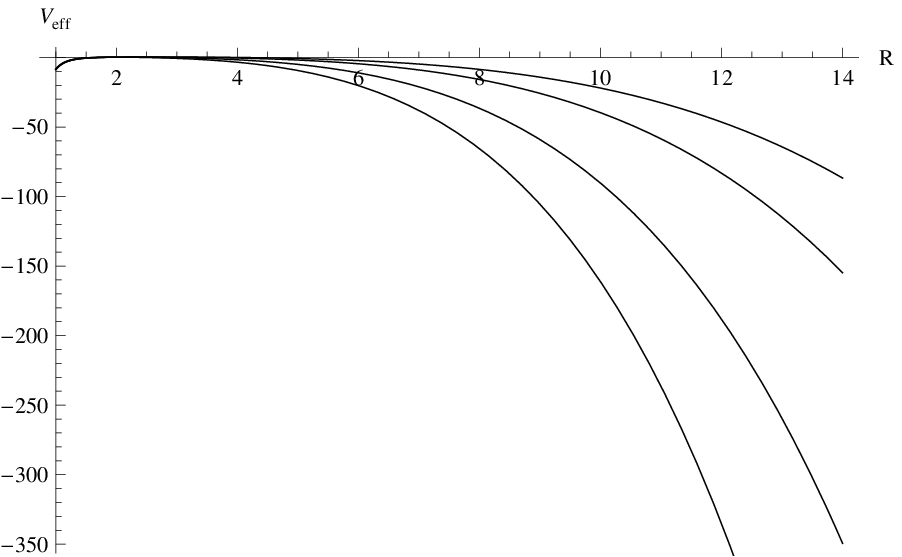, width=0.45\linewidth}\epsfig{file=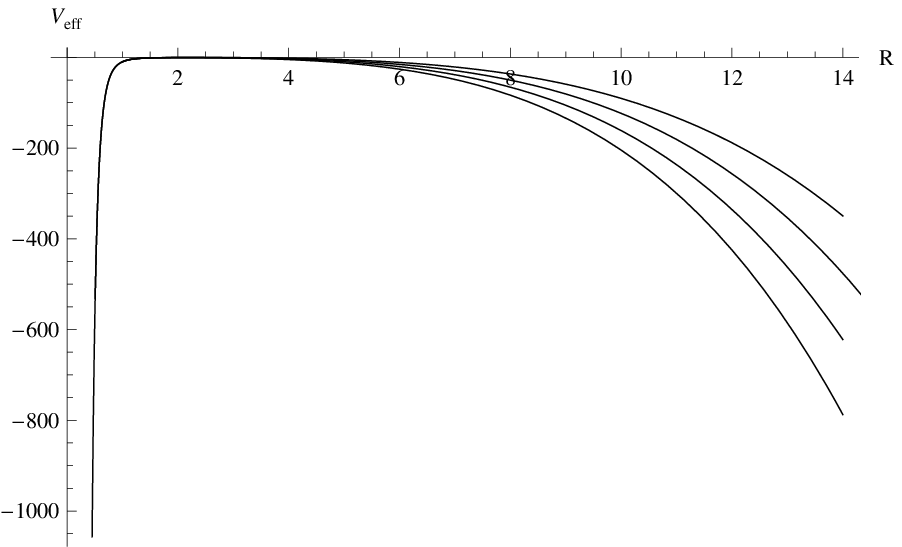,
width=0.45\linewidth} \caption{This figure describes the behavior of
effective potential (Eq.(\ref{34})). Both graphs correspond to
varying $M_+$ and $M_-$, keeping the remaining parameters fixed as
in previous cases.}
\end{figure}
\begin{figure}
\epsfig{file=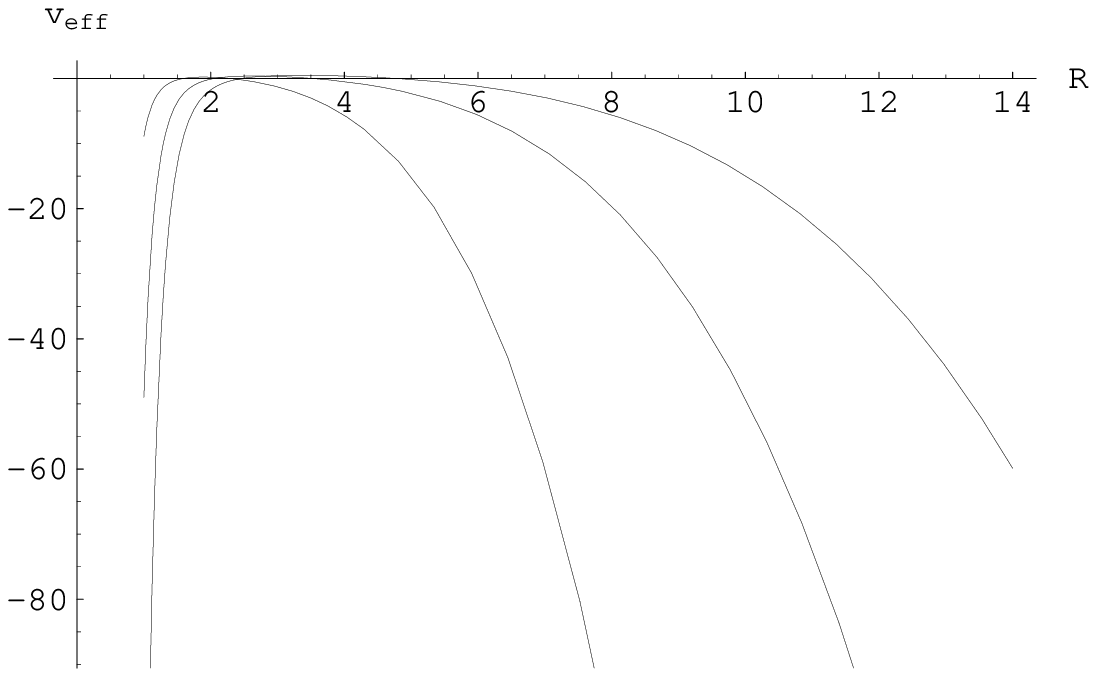, width=0.45\linewidth}\epsfig{file=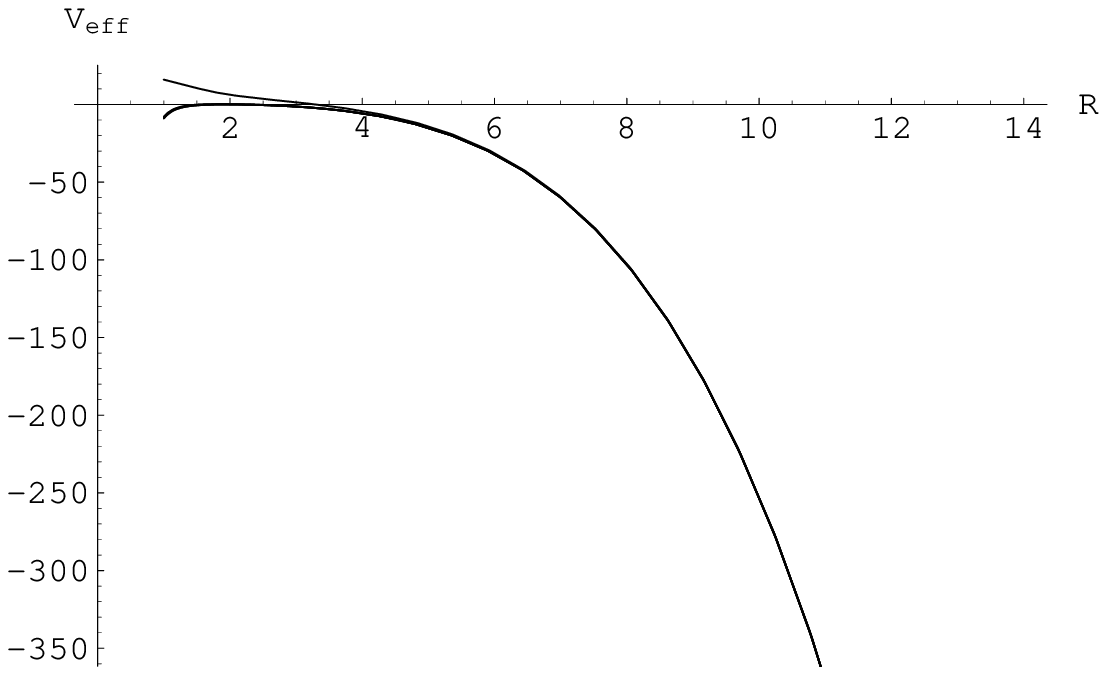,
width=0.45\linewidth} \caption{The left graph describes the
effective potential for massless scalar field with different values
of $\Omega$, keeping all the remaining parameters fixed as in
previous cases. The right graph represents the behavior of the
massless scalar field shell for different values of the charge $Q$.}
\end{figure}
We define the following two parameters:
$$[M]=M_+-M_-,\quad\overline{M}=\frac{M_++M_-}{2}.$$
Using these in Eq.(\ref{32}), it follows that
\begin{equation}\label{33}
\dot{R}^2+V_{eff}=0,
\end{equation}
where
\begin{equation}\label{34}
V_{eff}=1-\left(\frac{[M]}{2{\pi}{\Omega}^2}
\right)^2R^4+\left(\frac{Q}{R}\right)^2-\frac{2\overline {M}}{R}
-\frac{\pi^2{\Omega}^4}{R^{6}}.
\end{equation}

For the initial data of the shell, the left graph in Figure
\textbf{3} shows the increase and decrease in shell radius implying
the expansion and collapse of the massless scalar field shell. Thus
a massless scalar shell may expand or collapse depending on the sign
of velocity (i.e., $\dot{R}$) of the shell with respect to
stationary observer. The behavior of the potential depends on the
number of roots of the potential. If there is no root then the
scalar field shell either expands indefinitely or collapses to a
zero size from some finite value. If there is one non-degenerate
root then the shell expands to infinity or contracts to some finite
size. For one degenerate root, the shell will be in an unstable
equilibrium or collapses to form a black hole or naked singularity
\cite{26}.

The graphical representation of the effective potential with fixed
parameteric values of the model is shown in Figures \textbf{4, 5}.
Both graphs for varying $M_+$ and $M_-$ in Figure \textbf{4} and the
left graph in Figure \textbf{5} show that the effective potential
diverges for initial values of $R$ and then $V_{eff}\rightarrow
-\infty$ as $R\rightarrow\infty$. In these cases, the shell expands
to infinity or collapses to zero size. The right graph in Figure
\textbf{5} shows that the effective potential has one root and there
occurs unstable situation, after which potential diverges negatively
and shell expands or collapses. The cases in which collapse occurs,
the shell collapses to zero size by forming a curvature singularity
at which intrinsic Ricci scalar of the shell,
${R^{\mu}}_{\mu}=-\frac{2}{R^2}(2R\ddot{R}+{\dot{R}}^2)$, diverges.

\section{Massive Scalar Field}

In this case, we discuss the motion of a scalar field for which
potential term, $V(\phi)$ is determined by taking $p$ as an explicit
function of $R$. From Eq.(\ref{23}), we get
\begin{equation}\label{38}
\dot{\phi}^2=p+\rho,\quad V(\phi)=\frac{1}{2}(p-\rho).
\end{equation}
Also, from Eqs.(\ref{14}) and (\ref{15}), we get
\begin{equation}\label{38a}
\frac{d\rho}{d R}+\frac{2}{R}(p+\rho)=0.
\end{equation}
Here we use $p$ as an explicit function of $R$, \cite{25} i.e.,
$p=p_0e^{-kR}$, where $p_0$ and $k$ are constants. Inserting this
value of $p$ in Eq.(\ref{38a}), it follows that
\begin{figure}
\center\epsfig{file=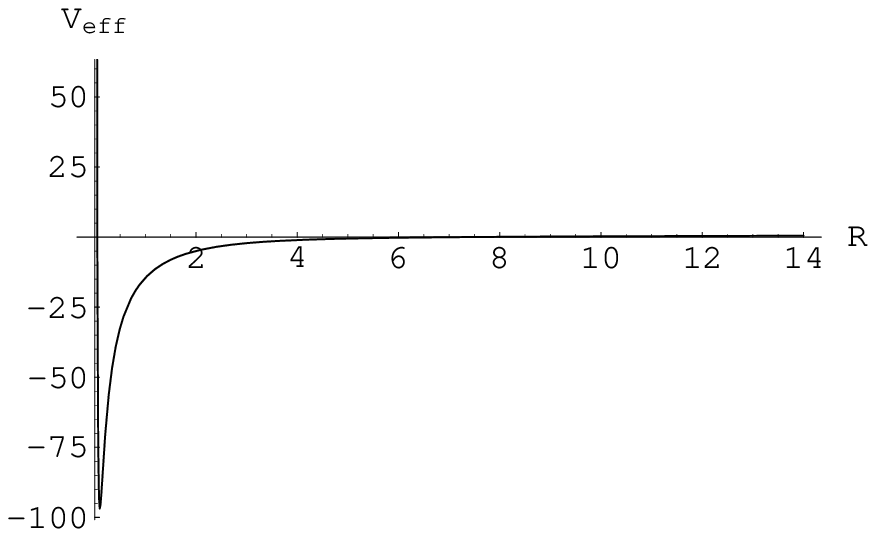, width=0.45\linewidth}
\epsfig{file=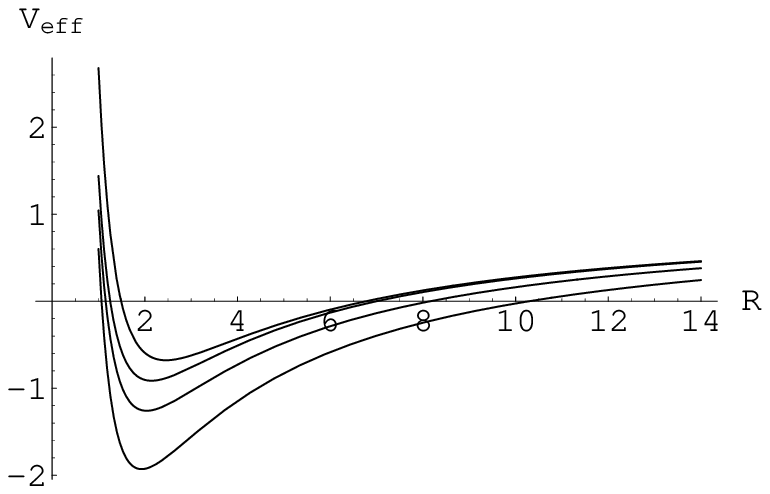, width=0.45\linewidth}\caption{The behavior of
effective potential for massive scalar field is shown for fixed
parameters as well with varying charge parameter.}
\end{figure}

\begin{equation}\label{40}
\rho=\frac{\chi}{R^2}+\frac{2(1+kR)p_0e^{-kR}}{k^2R^2},
\end{equation}
where $\chi$ is constant of integration. Notice that the above
equations satisfy the conservation equation (\ref{16}). Further,
applying the values of $p$ and $\rho$ in Eq.(\ref{38}), we get
\begin{eqnarray}\label{40}
V(\phi)&=&\frac{\chi}{2R^2}-\frac{p_0e^{-kR}}{2}\left(1-\frac{2(1+kR)}{k^2R^2}\right),\\\
{\dot{\phi}}^2&=&\frac{\chi}{R^2}+p_0e^{-kR}\left(1+\frac{2(1+kR)}{k^2R^2}\right).
\end{eqnarray}
These equations satisfy the KG equation (\ref{25}). Using
Eqs.(\ref{38})-(\ref{40}) in (\ref{26}), we have
\begin{equation}\label{41}
V_{eff}(R)=1-\left(\frac{M_+ -M_{-}}{m} \right)^2+\left(\frac{Q}
{R}\right)^2-\frac{(M_+ + M_{-})}{R}-\left(\frac{m}{2R}\right)^2,
\end{equation}
where
\begin{equation}\label{42}
m=4\pi R^2\rho\equiv 4\pi{\chi}+\frac{8\pi p_0e^{-kR}}{k^2}(1+kR).
\end{equation}

The behavior of effective potential for massive scalar field shell
is shown in Figure \textbf{6}. The left graph is effective potential
for massive scalar field (Eq.(\ref{41})) for $k=1,~\chi=3,p_0=1$ and
remaining parameters are fixed as in the massless scalar field case.
This implies that $V_{eff}\rightarrow -\infty$ as $R\rightarrow0$,
the massive shell collapses to zero size forming a curvature
singularity. The right graph represents effective potential for
massive scalar field shell for different values of $Q$. There appear
oscillations in the system. There exist such values of charge
parameter for which scalar field shell executes an oscillatory
motion. The oscillations occur at two points where {$V_{eff}$} cuts
the horizontal axis at more than one point. The values of $R$ for
which $V_{eff}=0$ are shown in right graph of Figure \textbf{6}
yielding zero velocity. This implies that the shell stops for a
moment and then expands or collapses. During the collapsing phase at
minimum values of the radius, the tangential pressure reaches its
maximum values while during the expansion, minimum pressure occurs
at maximum radius. In this way, scalar field shell performs the
oscillatory motion. The values of $R$ for which $V_{eff}=0$, and
intrinsic curvature of the shell is finite, are bouncing points
after bounce the shell either expands or collapse.

\section{Discussion}

In this paper, we have examined the dynamical behavior of the scalar
field thin shell. Using the Israel formalism, the equations of
motion have been formulated by taking the internal and external
regions to the boundary surface as RN solution. The equations of
motion are originally derived for perfect fluid and then are written
in terms of scalar field. The complete dynamics of the thin shell is
described by the equation of motion (\ref{17}) and the KG equation
(\ref{25}). The exact solution of these equations cannot be found,
however can be solved numerically. Firstly, we have solved these
equations by using the scalar field potential as quadratic
potential. This solution is shown in Figures \textbf{1, 2} which
represents the collapsing and expanding scalar field shell. It has
been found that scalar field decays out in the case of expansion
while it grows in the case of collapse.

To analyze further, we have taken massless scalar field and a
massive scalar field with $p$ as an explicit function of $R$. In the
case of massless scalar field, it has been found that the shell
radius is an increasing or decreasing function of the proper time
implying that the shell either collapses or expands. In this case,
the effective potential (Figures \textbf{3-5}) predicts that shell
can collapse to zero size by forming a curvature singularity or can
expand to infinity. For the massive scalar field case with $p$ as an
explicit function of $R$, we have evaluated the scalar field
potential instead of taking it as quadratic. It has been found that
the shell radius behaves like the massless scalar field and the
effective potential (Figure \textbf{6}) diverges negatively as the
radius of the shell approaches to zero and oscillating behavior is
noted in this case. This indicates that the shell collapses to zero
size forming a curvature singularity. There is also bouncing
behavior of the shell in this case. In the both (massless and
massive scalar field) cases, when shell collapses, the edge of the
shell coincides with the horizons of the interior black hole.

The results can be summarized as follows. We have found that there
are three possible phases (expanding, collapsing and oscillating
(bouncing)) during the dynamics of the scalar field in the present
configuration. When shell expands, it continues expanding forever
with constant velocity as the boundary surface is described by the
spatially homogenous spacetime. In case of collapse, a shell
collapses to zero size forming a curvature singularity at $R=0$,
where intrinsic Ricci scalar
${R^{\mu}}_{\mu}=-\frac{2}{R^2}(2R\ddot{R}+{\dot{R}}^2)$, diverges.
Also, the turning (bouncing) points occur when {$V_{eff}(R)=0$} at
more than one value of $R$. In this case, the oscillations occur
between two points where {$V_{eff}=0$}. The values of $R$ for which
$V_{eff}=0$ (right graph in Figure \textbf{6}), one gets zero
velocity. This implies that the shell stops suddenly and then
expands or collapses. During the collapse at $R=R_{min}$, the
tangential pressure reaches to its maximal values while during
expansion minimal pressure occurs $R=R_{max}$. In this way, the
scalar field shell performs the oscillatory motion.

It would be interesting to extend this work for more generic
geometry or using the polytropic equation of state to check the
validity of cosmic censorship hypothesis.

\vspace{0.25cm}

{\bf Acknowledgment}

\vspace{0.25cm}

We would like to thank the Higher Education Commission, Islamabad,
Pakistan for its financial support through the {\it Indigenous Ph.D.
5000 Fellowship Program Batch-IV}. Also, we highly appreciate the
fruitful comments of the anonymous referees.

\end{document}